\documentclass{emulateapj}
\usepackage{graphicx}
\usepackage{subfigure}
\usepackage{epsfig}
\usepackage{times}
\usepackage{natbib}
\usepackage{amsfonts}
\usepackage{amsmath}
\usepackage{amsbsy}
\usepackage{hyperref}
\usepackage[stable]{footmisc}
\usepackage{color}
\begin{document}
\newcommand{\project}[1]{\textsl{#1}}
\newcommand{\fermi}{\project{Fermi}}
\newcommand{\rxte}{\project{RXTE}}
\newcommand{\rhessi}{\project{RHESSI}}
\newcommand{\hz}{\,\mathrm{Hz}}

\title{Intermittency and Lifetime of the 625 Hz QPO in the 2004 Hyperflare from the Magnetar SGR 1806-20 as evidence for magnetic coupling between the crust and the core}
\author{Daniela Huppenkothen, Anna L. Watts}
\affil{Anton Pannekoek Institute for Astronomy, University of Amsterdam, Amsterdam  1098 XH,
  The Netherlands}
\email{d.huppenkothen@uva.nl}
\author{Yuri Levin}
\affil{Monash Center for Astrophysics and School of Physics, Monash University, Clayton, Victoria 3800, Australia}

\begin{abstract}
Quasi-periodic oscillations (QPOs) detected in the 2004 giant flare from SGR 1806-20 are often interpreted as global magneto-elastic oscillations of the neutron star. There is, however, a large discrepancy between theoretical models, which predict that the highest frequency oscillations should die out rapidly, and the observations, which suggested that the highest-frequency signals persisted for $\sim\!\! 100\,\mathrm{s}$ in X-ray data from two different spacecraft. This discrepancy is particularly important for the high-frequency QPO at $\sim\! 625 \hz$. However, previous analyses did not systematically test whether the signal could also be there in much shorter data segments, more consistent with the theoretical predictions. Here, we test for the presence of the high-frequency QPO at $625 \hz$ in data from both the \project{Rossi} X-ray Timing Explorer (\rxte) and the \project{Ramaty} High Energy Solar Spectroscopic Imager (\rhessi) systematically both in individual rotational cycles of the neutron star, as well as averaged over multiple successive rotational cycles at the same phase. We find that the QPO in the \rxte\ data is consistent with being only present in a single cycle, for a short duration of $\sim 0.5 \,\mathrm{s}$, whereas the \rhessi\ data are as consistent with a short-lived signal that appears and disappears as with a long-lived QPO. Taken together, this data provides evidence for strong magnetic interaction between the crust and the core.

\end{abstract} 

\section{Introduction}
\label{sec:introduction}
Asteroseismology is now firmly established as a precision technique for the study of stellar interiors. In this regard the detection of seismic vibrations from neutron stars was one of \rxte's most exciting discoveries, as neutron star seismology allows a unique view of the densest matter in the Universe. The vibrations, detectable as quasi-periodic oscillations (QPOs) in hard X-ray emission, were found in the tails of giant flares from two magnetars \citep{Israel05, Strohmayer05, Strohmayer06, Watts06}. Magnetars are highly magnetized neutron stars that exhibit regular gamma-ray bursts powered by decay of the strong magnetic field \citep{Thompson95}, and the rare giant flares thought to be associated with large-scale catastrophic magnetic field reconfiguration are apparently sufficiently energetic that they can set the entire star ringing. There is evidence for the presence of vibrations at both the same frequencies as observed in the giant flares, as well as previously unknown signals in the more frequent but less energetic smaller flares \citep{Huppenkothen13, Huppenkothen14}. It was realised immediately after their discovery that seismic vibrations from magnetars could constrain not only the interior field strength (which is hard to measure directly) but also the dense matter equation of state \citep{Samuelsson07,Watts07}. Over the last few years there has been intense development of seismic oscillation models that include the effects of the strong magnetic field, superfluidity, superconductivity, and crust composition \citep{Levin06,Levin07,Glampedakis06,Sotani2008,Andersson09,Steiner09,vanHoven11,vanHoven12, Colaiuda11,Gabler12, Gabler13, Passamonti13a, Passamonti13b,lee2008,asai2014}.   

The prevailing view is that QPOs, which have frequencies that lie in the range 18-1800 Hz, are associated with global magneto-elastic (most likely torsional/axial) oscillations of the star. The models have had some success in explaining the presence of long duration oscillations in the lower frequency band below 150 Hz; in axisymmetric models these oscillations are often associated with the turning points of the Alfv\'en continuum branches in the core \citep{Levin07}. However the higher frequency oscillations have proven to be something of a headache. Particularly problematic is a 625 Hz oscillation observed in data sets from two different satellites in the tail of the SGR 1806-20 giant flare \citep{Watts06, Strohmayer06}.  Frequencies in this range are predicted naturally in models where the crust vibrates independently without coupling to the core of the star, where they can be identified with the first radial overtone of the crustal shear modes \citep{Piro05}.  However for the field strengths expected (and measured) for magnetars, the motion of this crust mode should be strongly damped by converting its energy into that of the core Alfv\'en modes on timescales $\sim10-100 \, \mathrm{ms}$ \citep{Levin06,vanHoven11,Colaiuda11,Gabler12}. This would reduce surface amplitude and the signal should die out rapidly. The data analysis, by contrast, suggested that this signal persisted for $\sim\!\!\! 100 \, \mathrm{s}$ \citep{Strohmayer06}.

\begin{figure*}[htbp]
\begin{center}
\includegraphics[width=15cm]{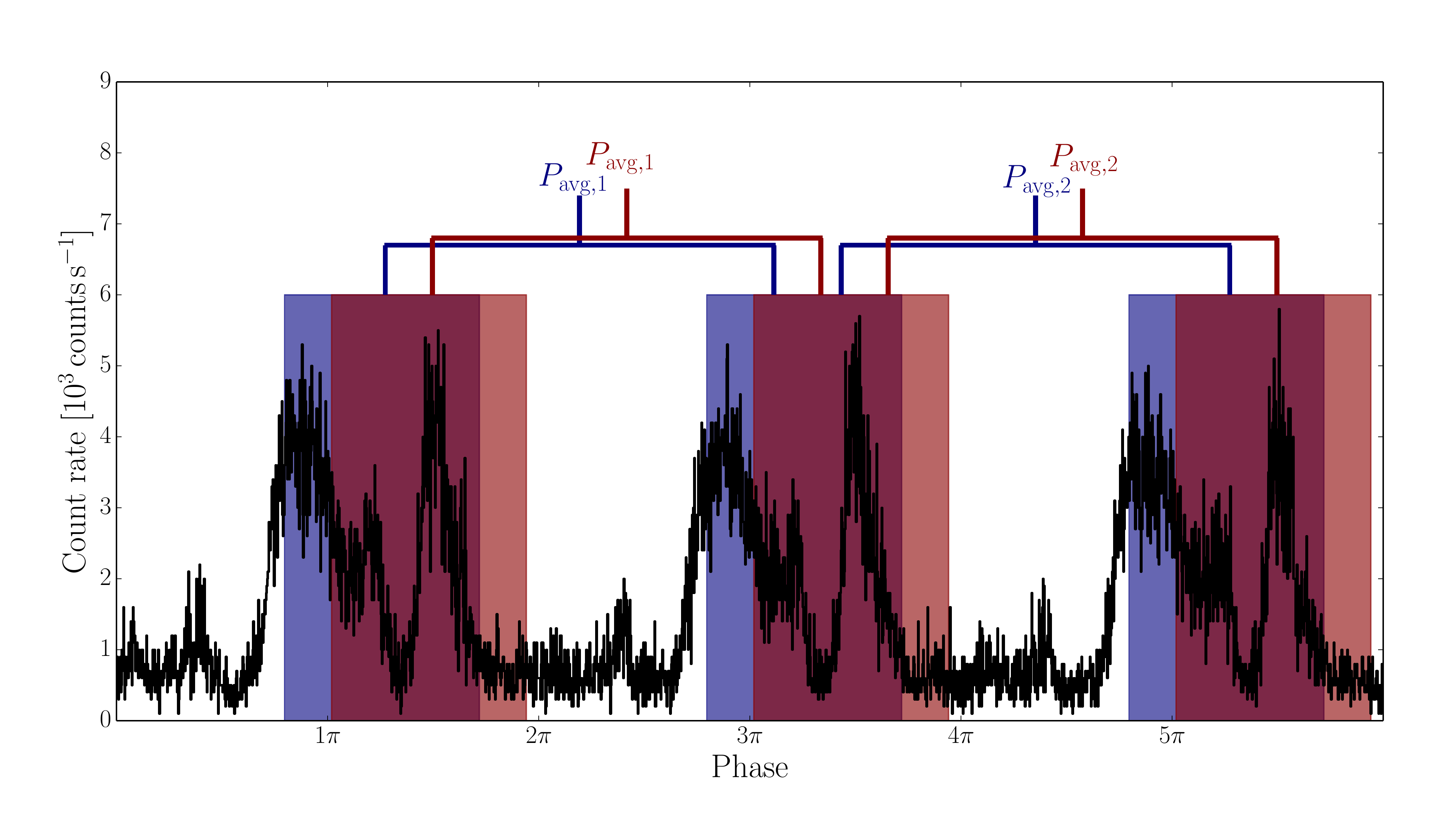}
\caption{Constructing averaged periodograms: in every rotational cycle of the neutron star, we select a stretch of length $t_{\mathrm{seg}}$ (the width of the shaded regions in this figure), starting $\delta t$ seconds apart (e.g. the distance between the start of the blue shaded region and the red shaded region). This way, we create $N_r$ overlapping segments per cycle, which start at the same
 rotational phase throughout the tail of the giant flare. Each segment is Fourier transformed into a Leahy-normalised periodogram; we then extract the power at $625 \hz$ (\rxte) or $626.5\hz$ (\rhessi). To search for long-lived signals, we averaged between $2$ and $10$ cycles (\rxte) and between $2$ and $19$ cycles (\rhessi) together in the following way. In order to average over two cycles, we extract the powers at the right frequency from all segments of cycles $1$ and $2$, and average powers together that match in phase (e.g. the two blue segments in this Figure). We repeat this procedure with cycles $2$ and $3$, and continue to the end of the flare. Similarly, we can average together cycles $1$, $2$ and $3$ to average over three cycles, followed by cycles $2$, $3$ and $4$, and so on. }
\label{fig:analysis1}
\end{center}
\end{figure*}

Various solutions to this problem have been explored. Attempts to explain it as an axisymmetric magnetically dominated oscillation associated with a turning point of the Alfv\'en continuum itself have proven difficult since at above $\sim\!\! 200\hz$ the branches of the Alfv\'en continuum are strongly overlapping in their frequency ranges.  The high-frequency crustal modes are therefore subject to resonant absorbtion by the continuum \citep{vanHoven11, vanHoven12}. The tangling of magnetic fields \citep{vanHoven11}  or coupling of the torsional (axial) oscillations to polar modes \citep{Colaiuda12,Lander10,Lander11} may break the continua and allow other types of oscillation to persist,
however none of the explored models showed oscillations at frequencies as high as $600$Hz. Taking into account effects associated with superfluidity may also be the answer. Superfluidity can move the continua such that damping is reduced \citep{vanHoven08, Andersson09, Passamonti13a} and may result in resonances between crust and core that could prolong mode lifetimes \citep{Gabler13, Passamonti13b}.  In fact in their axisymmetric model \citep{Gabler13} found an oscillation at $\sim 600$Hz; however, this work
did not demonstrate  numerical convergence of this result. Analogous unpublished numerical experiments
by van Hoven, albeit with an entirely different method, have shown similar oscillation that featured an amplitude that was decreasing as a function of numerical resolution.
It is clear that the time-dependence of the amplitude of the $625$Hz QPO is needed
to constrain the theoretical models.

In this paper we revisit the data analysis method, which as it turns out was not well-suited to address the question of whether the properties of the 625 Hz signal are consistent with rapid decay on very short timescales.
To understand why, we need to review the data analysis procedures that were followed. The amplitude of the strongest signal found during the SGR 1806-20 giant flare, at 92 Hz, was found to be strongly dependent on rotational phase. The signals at other frequencies were then identified by taking short segments (typically $30\%$ of a rotational cycle, which corresponds to 2.3 s for SGR 1806-20), of consecutive rotational cycles, and averaging together power spectra from these individual segments. Significance was estimated using standard procedures for averaged power spectra \citep{vanderKlis89}, with corrections for the deviation of noise powers from a pure Poisson distribution, particularly at low frequencies. Having identified a significant signal (as compared to the null hypothesis), start/end points and hence durations for a signal of a given frequency were estimated by adding or subtracting power spectra from segments of rotational cycles at the ends of sequence, and identifying the set for which significance was maximised. This method was adequate to identify signals that were significant compared to the null hypothesis. However it does not distinguish between a signal that is present at a constant low level throughout the relevant segment of every rotational cycle, and one that is present for a much shorter time in perhaps only a few non-consecutive rotational cycles in the sequence.
\begin{figure}[h!tbp]
\begin{center}
\includegraphics[width=9cm]{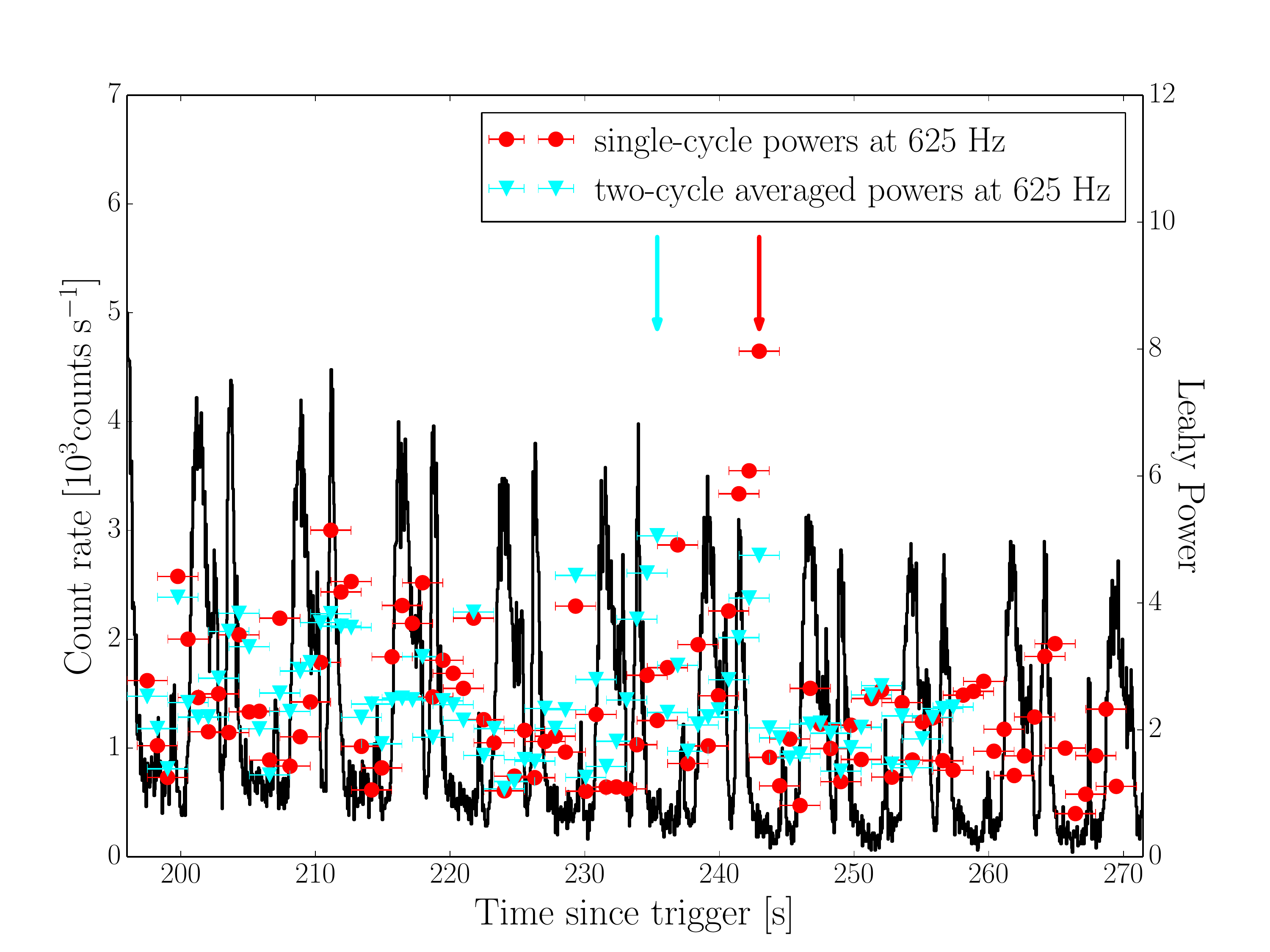}
\caption{The test statistic we use is defined as the maximum power over all segments in the giant flare. Here, an example for the \rxte\ observation: we extract powers at $625 \hz$ in $3\mathrm{s}$ segments starting $\sim 0.75 \mathrm{s}$ apart, and bin over $2.66 \hz$. The resulting Leahy-normalised powers (red) are plotted on top of the \rxte\ light curve (black). The x-axis error bars denote the width of each segment. The maximum power (red arrow) is found in a segment on the falling edge of the pulse, in a cycle $\sim 240 \mathrm{s}$ after the trigger. In order to assess the significance of this power, we repeat the analysis described in the text on simulated giant flare light curves without the presence of a QPO. As for the data, we extract the maximum power over all cycles and segments, and compare the distribution of these maximum powers to that observed in order to compute a p-value. This ensures that the number of trials for the (non-independent) segments is correctly accounted for. For comparison, we also plot 2-cycle averaged powers as described in the text and Figure \ref{fig:analysis1} (cyan symbols). Powers averaged over two cycles are generally lower, but this is expected: as powers are averaged, the noise distribution narrows as well. As for the non-averaged powers, we extract the maximum power over all cycles and segments as the relevant test statistic (cyan arrow). Note that the maximum power is shifted to a segment at the same phase as the segment with the maximum non-averaged power, but a cycle earlier; also note that there is necessarily no 2-cycle averages for the last cycle. }
\label{fig:teststatistic}
\end{center}
\end{figure}
\begin{figure*}[htbp]
\begin{center}
\includegraphics[width=18cm]{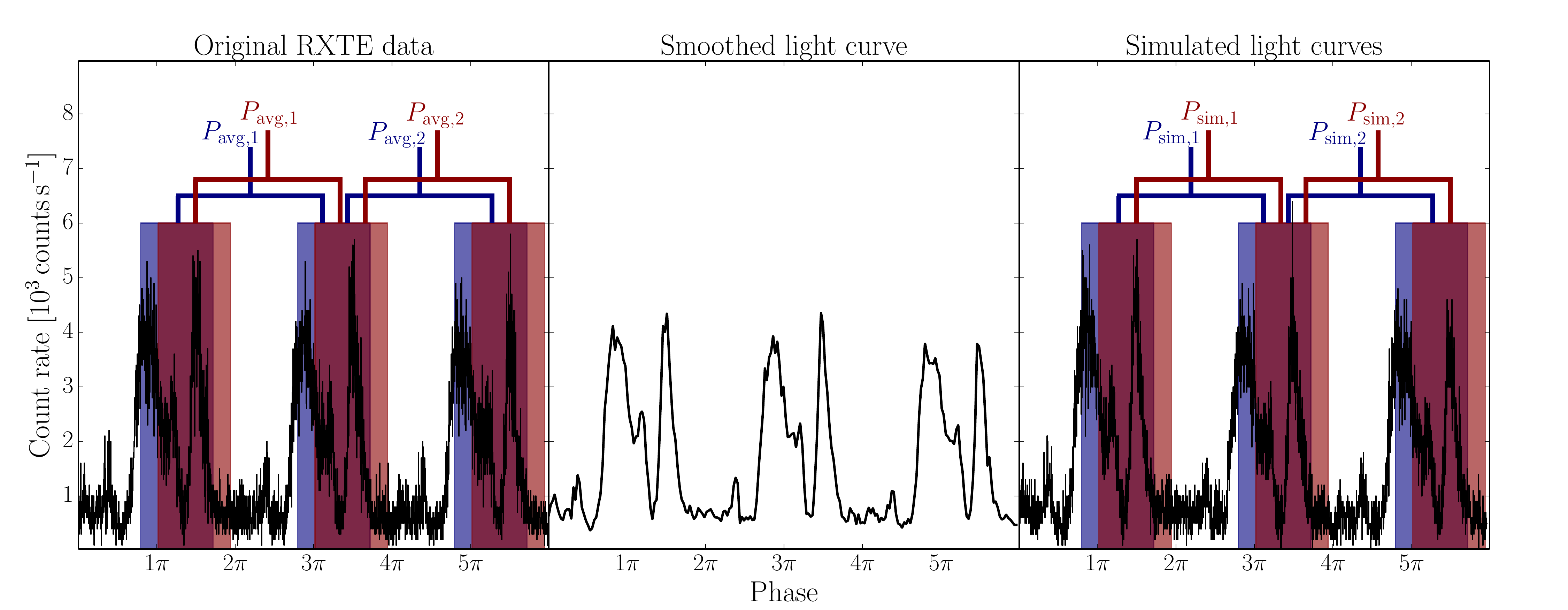}
\caption{Overview of the analysis procedure (here shown is the \rxte\ light curve). Left panel: first, we extract powers at $625\hz$ (\rxte) and $626.5\hz$ (\rhessi) from individual, overlapping segments, and average powers from between $2$ and $9$ (\rxte; $2$ and $19$ for \rhessi) cycles as described in the text and shown in Figure \ref{fig:analysis1}. In order to test for significance, we create simulations from the original data in the following way. We smooth the light curve to $0.01\,\mathrm{s}$ in order to remove all traces of the high-frequency QPO (middle panel; smoothing exaggerated for illustrative purposes). We then sample the light curve at the original, high time resolution and produce $N_\mathrm{sim}$ realisations by introducing Poisson noise to simulate the effects of a photon counting detector. We can then perform the same procedure as on the observed data on our simulations (right panel), and test the observed (averaged) powers against distributions of simulated powers with the QPO removed.}
\label{fig:analysis2}
\end{center}
\end{figure*}

We would like to test the specific question of whether the data are consistent with a model where whenever the 625 Hz signal appears, it dies out on a timescale that is much shorter than the segment durations considered in the previous analysis. We allow the possibility that the signal may be excited several times during the tail of the giant flares (perhaps by aftershocks)\footnote{The possibility of excitation late in the tail of the flare is already supported by the fact that the strongest 92 Hz signal does not appear until about $100\,\mathrm{s}$ into the tail.}. A secondary goal is therefore to determine how many times, and at what level, such a signal must be excited to be consistent with the data, if the data is of a high enough signal-to-noise ratio to determine such an effect. In this paper we therefore develop a more sophisticated analysis method that is tailored to address the specific question of whether the data are consistent with rapid die-out of the 625 Hz signal, and the conditions that must be met in terms of re-excitation for this to be the case. It is interesting to note that the possibility that the data might be consistent with a sequence of rapidly decaying pulses may explain their apparent coherence. The width of many of the signals, including the 625 Hz QPO, is consistent with what one would expect for an exponentially decaying but strictly periodic signal with a decay timescale shorter than 1s. The fact that this was inconsistent with the apparent durations was noted by \citet{Watts11}, and taken as evidence that the signals were genuinely quasi-, rather than strictly, periodic.

\section{Data Analysis \protect\footnotemark[2]} 
\label{sec:analysis}

We include data sets from two different space telescopes in our analysis: The \project{Rossi} X-ray timing Explorer (\rxte), and the \project{Ramaty} High Energy Solar Spectroscopic Imager (\rhessi). An overview of the \rxte\ data is given in \citet{Israel05}. Data were recorded in $\mathrm{Goodxenon\_2s}$ mode, allowing for time resolution up to $1 \, \mu \mathrm{s}$, high enough to study high-frequency QPOs, in an energy range between $4 \, \mathrm{keV}$ and $90 \, \mathrm{keV}$.
Observations taken with \rhessi\ are detailed in \citet{Watts06}. Following their analysis, we only used photons recorded with the eight front segments of the telescope: The rear segments recorded not only direct photons but also a bright component reflected from the Earth. The delay between the two smears out the signal and precludes searches for high frequency signals using data from the rear segments. The high-frequency QPO in the \rhessi\ data is only seen in the energy range between $100 \, \mathrm{keV}$ and $200 \, \mathrm{keV}$, where \rxte\ cannot observe, but not at the lower energies. We hence filter for the $100 - 200 \, \mathrm{keV}$ energy band.  \rhessi\ has a comparable native time resolution to \rxte: $1$ binary $\mu\mathrm{s}$ ($2^{-20} \, \mathrm{s}$). All data are barycentered, that is, corrected for the motion of the space craft through space to avoid systematic effects in the timing analysis.
\footnotetext[2]{All relevant data products and analysis scripts used for the analysis below, to produce all results and Figures shown in this paper, are available for public download at \hyperref[https://github.com/dhuppenkothen/giantflare-paper]{https://github.com/dhuppenkothen/giantflare-paper}.}

Note that the QPO was not detected in the data from both satellites at the same time: it appeared first $\sim\!\! 80\,\mathrm{s}$ after the initial spike of the giant flare in the high energies seen in \rhessi, and later, $\sim\!\! 196 \,\mathrm{s}$ after the initial spike at lower energies observed in \rxte.
For the \rxte\ data, we concentrated on the part of the light curve where the $625 \, \mathrm{Hz}$ QPO was originally found, from around $190\, \mathrm{s}$ after the onset of the flare to the end of the observation. This encompasses a total of $15$ rotational cycles of the neutron star. The \rhessi\ observations place the same QPO at a slightly different frequency ($626.5 \, \mathrm{Hz}$ as opposed to $625.5 \, \mathrm{Hz}$). For the latter, we search the range from $80\, \mathrm{s}$ to $225 \, \mathrm{s}$ from the onset of the flare, or equivalently 19 cycles.

In the original analyses of both data sets, the QPO was detected in phase-resolved periodograms averaged over a large number of cycles, but it has never been clear whether the data require that the QPO is present consistently over this large number of cycles, or whether there may be a few strong, re-excited QPOs scattered over the entire period where the QPO was observed. In averaged periodograms, both would look very similar.
In order to see whether the data would support an alternative explanation - strong, re-excited signals - we test systematically for the presence of a strong QPO in both data sets against the simple null hypothesis (no QPO) cycle by cycle, as well as for averaged periodograms while varying the number of cycles per averaged periodogram. If there is indeed a signal present in only a few cycles, and the data are of high enough quality to clearly detect them, this analysis will be able to both quantify the significance of the detected signals, as well as their location in time in the tail of the giant flare.

To search for QPOs, we split each rotational cycle into a number $N_\mathrm{r}$ of overlapping segments of length $t_{\mathrm{seg}}$, starting at intervals of $\Delta t$ seconds apart (see Figure\ref{fig:analysis1} for an illustration). For each of these segments, we binned the event data to a time resolution of $\delta t = 5 \times10^{-4} \, \mathrm{s}$ (equivalent to a maximum frequency of $1000 \, \mathrm{Hz}$), computed the periodogram and extracted the power at the frequency where the QPO was observed. For each periodogram we tested the significance of the maximum observed power over all segments and cycles against $N_{\mathrm{sim}}$ simulations of the null hypothesis (no QPO), which are constructed in the following way (see also Figures \ref{fig:teststatistic} and \ref{fig:analysis2} for an illustration).

As a first step, we smoothed out the light curve to a resolution of $0.01 \, \mathrm{s}$, or equivalently $100 \, \mathrm{Hz}$, ensuring that all possible variability at smaller time scales is eliminated from the data. We then interpolated back to the original time resolution used ($\delta t = 5\times10^{-4} \, \mathrm{s}$), and added Poisson noise to this smoothed light curve $N_{\mathrm{sim}}$ times. This represents the null hypothesis that the QPO is not present, and that any variability measured at $625 \, \mathrm{Hz}$ is solely due to photon counting noise in the detector. For each of our $N_{\mathrm{sim}}$ simulations, we performed exactly the same analysis as for the observed data. We can then compare the real powers we measured for a given segment to the distribution of simulated powers in that segment. Additionally, we can compare the maximum power observed at $625 \, \mathrm{Hz}$ ($626.5 \, \mathrm{Hz}$ for the \rhessi\ data) for all segments in our observed data for the maximum powers at this frequency in the ensemble of simulated light curves. A formal comparison between observation and data is done by using the simulated powers at the QPO frequency to construct a probability distribution for the power at this frequency under the null hypothesis. Our confidence in the observed power being an outlier under the null hypothesis is expressed as the integrated probability of finding a power at least as high or higher than the observed value, also known as the p-value. A smaller p-value corresponds to a smaller probability of making the observation under the null hypothesis. Under the assumption of a Gaussian distribution, one can directly translate p-values to the more commonly used $\sigma$-values for significance. 

While in principle, the probability distribution for periodograms consisting mainly of Poisson noise is very well known, there are two reasons why the simulations detailed above are necessary: (1) the data do not consist purely of Poisson noise; while we do not expect any significant contribution from e.g. the pulse profile at these frequencies, we cannot exclude it, either. Additionally, the segments we choose at different phase intervals have vastly different overall shapes, including long-term trends. Constructing simulations in the way we described above ensures that these effects are taken into account, without having to know them in detail. (2) When performing this analysis on overlapping segments, the individual powers extracted at $625 \, \mathrm{Hz}$ are not independent from segment to segment. This leaves doubt about the necessary correction for the number of segments searched (as it becomes more likely to see a high power simply by chance when searching a large number of periodograms). One can make the most conservative assumption: completely independent trials for each segment, but this is likely too conservative and unnecessarily constrains our predictive power. By performing the simulations in the described way, and searching over all segments, the number of trials is already taken into account in the correct way, allowing us to quote accurate p-values while not being overly conservative.

In addition to testing all segments individually, we constructed phase-averaged periodograms in the following way. We match all periodograms belonging to segments that start at the same phase with each other. In order to construct the two-cycle average, we average the same phase bins for two consecutive cycles together, and again extract the power at the relevant frequency (see also Figure \ref{fig:analysis1}). We then do the same for the next cycle, and so on until we reach the end of the data under consideration. The result is a moving average over subsequent rotational cycles, where the averaged periodograms match in phase.
Similarly, we can construct three-cycle averages by combining periodograms from three consecutive cycles, and so forth, until we average the maximum number of cycles in our particular data set.  
Note that powers in averaged periodograms are not independent of either neighbouring segments or phase-matched segments, since each power is averaged at least twice with either neighbour (or more times in the case of constructing averaged periodograms from a larger number of rotational cycles). 
In each case, we construct simulations in the same way as detailed above by smoothing the light curve to compare against the null hypothesis, and construct phase-averaged periodograms in the same way as for the data for each of our $N_\mathrm{sim}$ simulations. Consequently, we can construct simple p-values for the significance of the maximum power over all cycles and segments in the averaged periodograms. 

\begin{figure*}[htbp]
\begin{center}
\includegraphics[width=18cm]{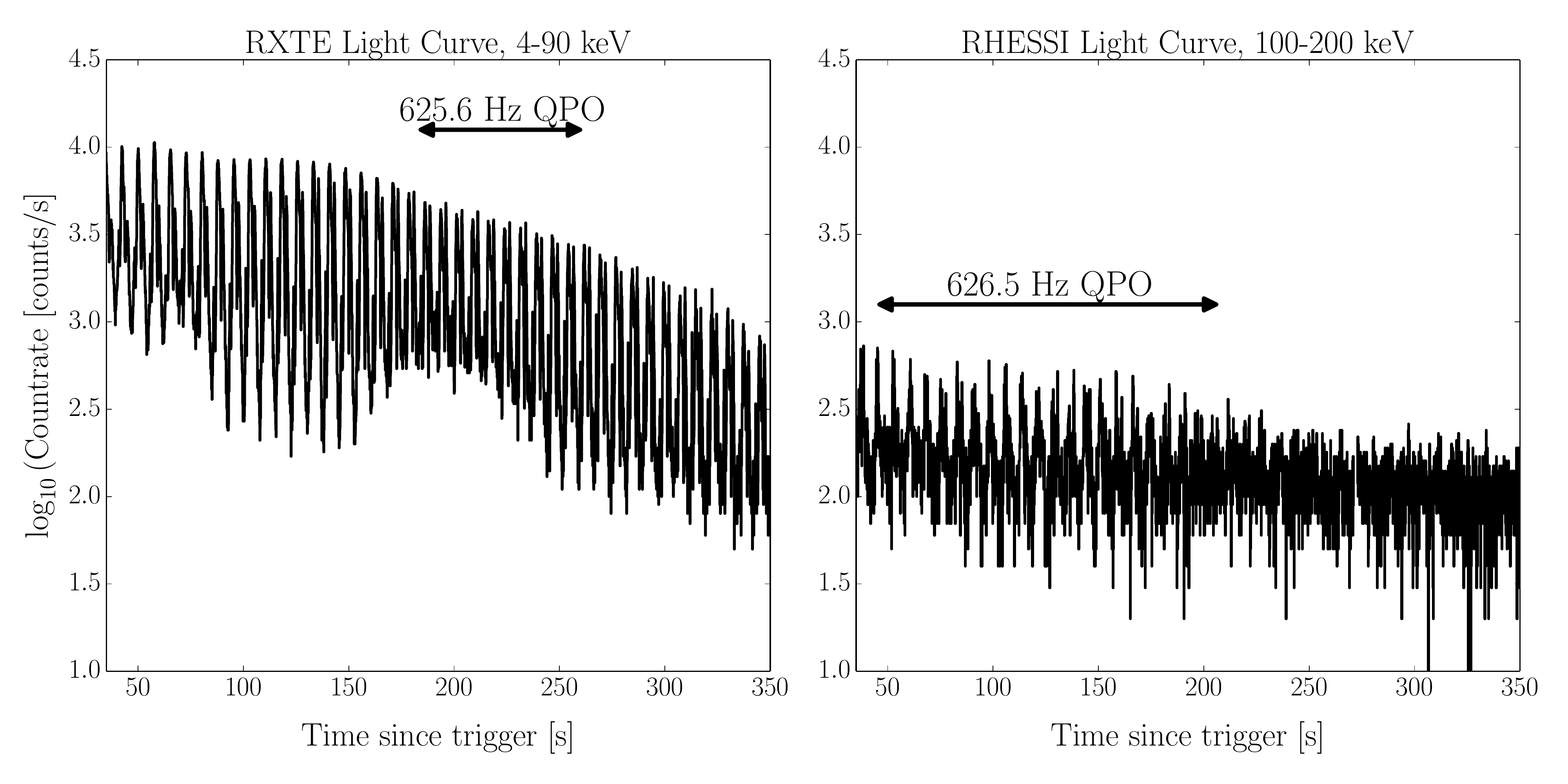}
\caption{Light curves of both the \rxte\ data (left panel) and the \rhessi\ data (right panel) in the relevant energy ranges. Note that both light curves are plotted on the same (logarithmic) scale, showing the vastly different data quality between the two instruments. 
The cycles in which the QPO was previously detected by \citet{Strohmayer06} (for the \rxte\ data, left panel) and \citet{Watts06} (for the \rhessi\ data, right panel) are indicated by horizontal arrows.}
\label{fig:lcs}
\end{center}
\end{figure*}

\begin{figure}[htbp]
\begin{center}
\includegraphics[width=9cm]{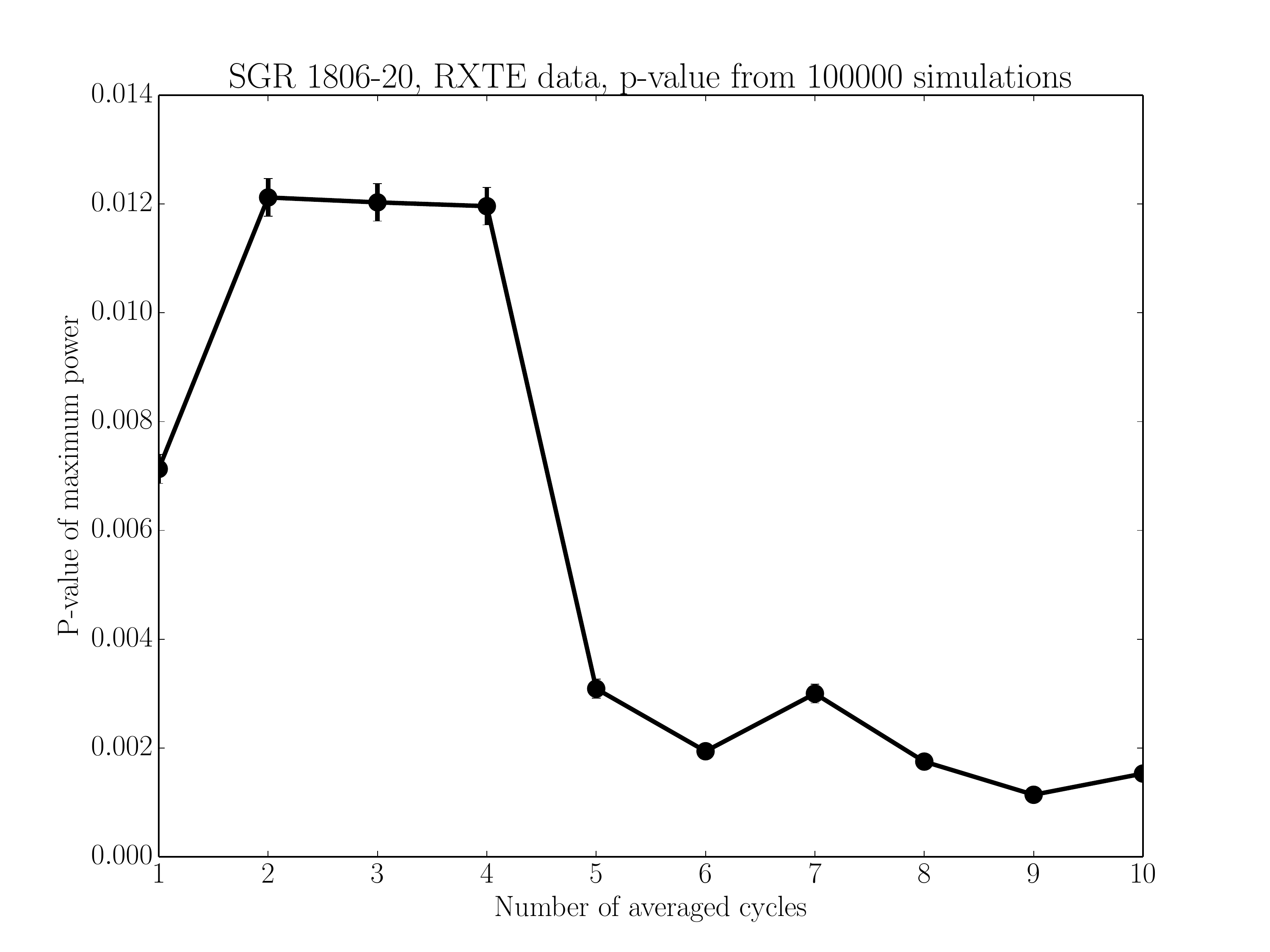}
\caption{Integrated probability of observing a signal at $625 \, \mathrm{Hz}$ in the \rxte\ data at least as high or higher than the observed power as a function of the number of cycles averaged. For each data point, we extracted the maximum power from all (averaged) segments over the entire length of the searched part of the giant flare light curve, for both the observation and the simulated light curves. A smaller p-value corresponds to a higher significance: There is a significant signal in a single cycle, starting at $t = t_0 + 239 \, \mathrm{s}$, where $t_0$ is the trigger of the observation, as well as significant detections when averaging five consecutive cycles or more. Errors on the p-values are derived from the theoretical approximation valid for small probabilities $\Delta p = \sqrt{p (1-p)/N}$, where $p$ is the p-value and $N_\mathrm{sim}$ is the number of simulations. Note that this likely underestimates the real error on the p-value, as it excludes any systematic effects.}
\label{fig:rxte_pvals}
\end{center}
\end{figure}

The simulations detailed above will only give us an idea of whether the data are consistent with the null hypothesis. In order to test more complicated hypotheses, for example the presence of the QPO in specific cycles, at specific phases, or lasting for a specific duration with a particular amplitude, we can inject an artificial sinusoidal signal with a given frequency, duration and amplitude into a smoothed light curve and regard this as our new data set, to be analysed in exactly the same way as the real observations, such that for each such light curve, we get p-values for the strongest signal as a function of the number of cycles averaged. Depending on our knowledge of the system and the conclusions drawn from the data, the parameter space for these simulations can be very large: we can vary the number of cycles into which the signal is injected, the exact sequence of cycles in which the signal is injected, the phase at which the QPO is observed, the fractional rms amplitude of the signal, and the duration of the QPO within a single cycle.

All artificial signals we inject are at the same frequency as the observed signal and at a constant fractional rms amplitude, i.e. their absolute amplitude varies with the pulse profile. For each simulated light curve, the starting phase of the signal is randomised.

\section{Results}
\label{sec:results}

\subsection{RXTE}
\label{sec:rxte_results}

\citealt{Strohmayer06} reported a detection of a strong QPO when averaging nine consecutive cycles to a significance of $p < 1.1 \times 10^{-7}$ single-trial, or $10^{-5}$ trial-corrected, and a fractional rms amplitude of $8.5\%$ (see Figure \ref{fig:lcs} for light curves and arrows indicating the part of the giant flare where the QPO was detected in both the \rxte\ and \rhessi\ data). These values are based on comparing the observed power to a theoretical Poisson distribution after dividing out a model fit for the low-frequency powers. They also report the detection of the same feature in an averaged periodogram of specific two cycles from the segment where the QPO was found originally with $p < 1.1 \times 10^{-6}$ single-trial and a fractional rms amplitude of $18.3\%$. A third averaged periodogram six cycles before the previous one showed a signal to $p < 4.4 \times 10^{-6}$ (single-trial), but no trial-corrected p-value was calculated for either of the last two reported detections, so estimating the actual significance of these latter two signals, as compared to the nine-cycle average, is impossible.

We first repeated the analysis from \citealt{Strohmayer06} in order to reproduce their results, paying special attention to the overall number of cycles during which the signal was present, as well as the duration of the presence of the QPO in any individual cycle.
We searched individual segments of $t_{\mathrm{seg}} = 3 \, \mathrm{s}$ length, starting every $\approx 0.75 \, \mathrm{s}$, such that consecutive segments overlap by $2.25 \, \mathrm{s}$. We extracted the power at $625 \, \mathrm{Hz}$, then ran $10^5$ simulations as described in the previous section and compared the powers at $625 \, \mathrm{Hz}$ for each segment as well as phase-averaged segments to the powers extracted in the same way from the simulations.

In order to answer the question whether the QPO is present in all nine cycles, as reported in \citet{Strohmayer06}, or whether it is only present in a single cycle, we constructed phase-averaged periodograms averaging up to nine periodograms, and constructed p-values as described in Section \ref{sec:analysis}. Figure \ref{fig:rxte_pvals} presents the results: the significance of the QPO depends strongly on how many cycles are averaged. The smallest p-value, corresponding to the smallest probability that the observed power could be due to a chance occurrence, occurs when averaging 9 cycles, consistent with the results in \citet{Strohmayer06}. However, we also note that for a signal consistently present over all 9 cycles, we would expect the p-value to slowly drop with an increasing number of averaged cycles. This is not observed: there is a significant detection in a single cycle, starting at $t_0 + 239 \,\mathrm{s}$ (where $t_0$ denotes the time when \rxte\ triggered on the giant flare emission), with a significance of $p = 0.007$, trial-corrected\footnote{Note that our extracted powers, and consequently the resulting significances, do not exactly match those of \citet{Strohmayer06}. This is largely due to an error in the reported channels in that paper ($6-190$ as opposed to $10-200$, as reported in the paper; Strohmayer, private communication), but the difference is small and has no bearing on the qualitative results of this work.}. When averaging neighbouring cycles into that strong signal, the p-value first increases by a factor of two, then drops sharply for an average of 5 cycles. This could perhaps indicate that the QPO is only intermittently present. In order to test this hypothesis, we constructed the periodogram for the eight cycles, \textit{excluding} the cycle starting at $t_0 + 239\, \mathrm{s}$, where the strongest signal is detected, and compared this to the simulated light curves as well. In this case, the p-value for detecting a QPO in this phase bin drops to insignificance, $p = 0.13$. This is a clear sign that either the QPO is confined to just one cycle, or else that the signal is buried underneath the noise for the other cycles. 

\begin{figure}[htbp]
\begin{center}
\includegraphics[width=9cm]{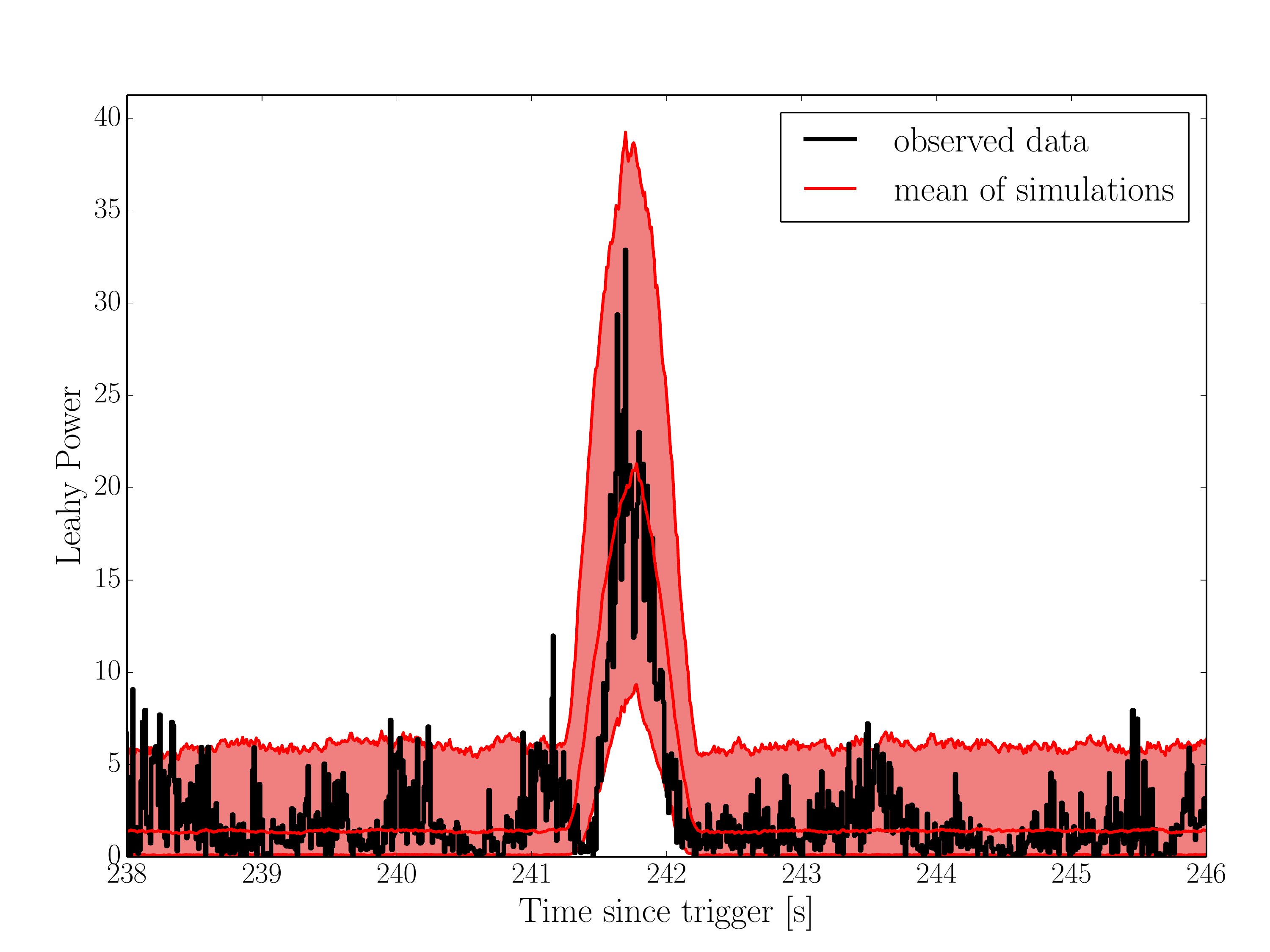}
\caption{The cycle of the giant flare, observed with \rxte, with the strongest QPO signal. Each data point (in black) corresponds to the power at $625 \,\mathrm{Hz}$ extracted from a $3\,\mathrm{s}$ segment starting $1.5\,\mathrm{s}$ before the time stamp of the data point. Each segment is separated by $0.01\,\mathrm{s}$, and thus neighbouring segments overlap by $2.99 \,\mathrm{s}$. Note the sharp rise in power at $\approx t_0+ 241.5\,\mathrm{s}$: as the sliding window shifts into the part of the part of the light curve where the QPO appears, the power at $625\,\mathrm{Hz}$ rises sharply as well. In order to constrain the duration and amplitude of the signal, we simulated $1000$ light curves, where we first smoothed out all variability above $100\,\mathrm{Hz}$, then added a single sinusoidal signal of $0.5\,\mathrm{s}$ duration and a fractional rms amplitude of $0.22$, starting at $241.5 + t_0\,\mathrm{s}$, and analysed these light curves in exactly the same way as the data. The thick red line corresponds to the mean of these $1000$ simulations for a given segment, in analogy to the data (in black). The shaded area constrains the $5\%$ and $95\%$ quantile ranges derived from the simulations, indicating that the observed powers are well-represented by our simulations.}
\label{fig:rxte_sims}
\end{center}
\end{figure}

We note that our p-values are generally higher (denoting lower significance) than those reported in \citet{Strohmayer06}. This is largely due to a combination of the way we have taken into account the number of trials (simulating the entire analysis on fake data without the QPO ensures that the number of trials is correctly taken into account) as well as the fact that for the simulated light curves, the distribution of powers at 625 Hz does not strictly follow the expected theoretical $\chi^2$-distribution with $16n_{\mathrm{cycles}}$ degrees of freedom (the number of degrees of freedom corresponds to twice the number of frequencies averaged times the number of cycles averaged to obtain a given power).
We also note that we fail to reproduce the marginal detection of a QPO at the same frequency six cycles before the cycle with the strongest incarnation of the $625\, \mathrm{Hz}$ QPO. 

Given that the data are far more consistent with a signal being present in only one cycle than in a longer stretch of data, we attempted to constrain the width of the QPO in the strongest cycle. First, we repeated the analysis described above, but with time segments of $t_{\mathrm{seg}} = 3 \, \mathrm{s}$ duration, starting every $0.01\,\mathrm{s}$ apart, effectively providing a finely resolved sliding window over the cycle where the QPO is strongest. If one then plots the strength of the signal with time (see Figure \ref{fig:rxte_sims}), one can track the strength of the QPO over the course of the star's rotational cycle. As more signal is included in a given segment, the power will rise, until the entire QPO is included. Similarly, as the sliding window moves out of the time frame where the QPO is located, less and less signal is included, and the power drops. We show the resulting plot in Figure \ref{fig:rxte_sims}. Similarly to Figure 3 in \citet{Strohmayer06}, the QPO seems to be present only for a short period of time. 

We introduced an artificial sinusoidal signal into a single cycle in 1000 simulations, in the same part of the rotational cycle as the real QPO, for a duration of $0.5\, \mathrm{s}$ and a fractional rms amplitude of $0.22$. The amplitude of the sinusoidal signal varies with the underlying giant flare emission, such that the fractional rms amplitude remains constant. In Figure \ref{fig:rxte_sims}, we show the mean power for each segment out of $1000$ simulations, as well as the 5\% and 95\% quantiles derived from these simulations, compared with the powers derived from the real data. The observed powers are easily reproduced with a short signal of $0.5\, \mathrm{s}$ duration. We note that $0.5\,\mathrm{s}$ is only a small fraction of the rotational period of the neutron star itself \citep[$7.5477 \,\mathrm{s}$,][]{Woods07}, thus cannot be easily explained by the region of the neutron star affected by the oscillations moving in and out of the line of sight.
\begin{figure}[htbp]
\begin{center}
\includegraphics[width=9cm]{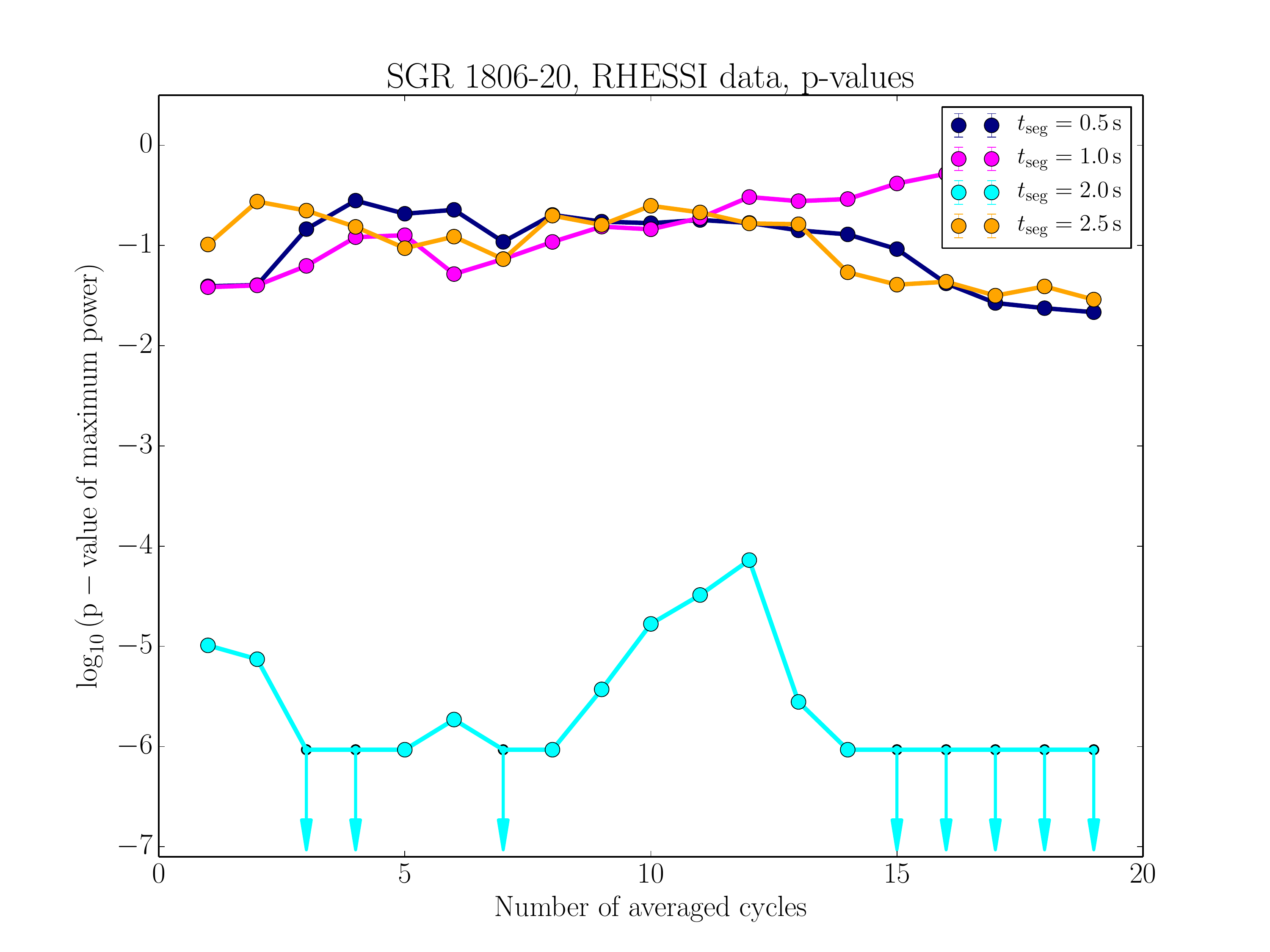}
\caption{\rhessi\ data: p-values for measuring a power at $626.5\, \mathrm{Hz}$ in any of the segments at least as high or higher than the ensemble of powers derived from all segments from $N_{\mathrm{sim}}$ simulations. We chose $N_\mathrm{sim}=10^{4}$ for segments of length $t_{\mathrm{seg}} = (0.5, 1.0, 2.5) \,\mathrm{s}$. Because the QPO is much more significant in the $2\,\mathrm{s}$ long segments, we ran a total of $N_{\mathrm{sim}}=10^{6}$ in this case, at which point we exceeded computational feasibility. P-values smaller than $10^{-6}$ are indicated by arrows as upper limits. We show all p-values as a function of the number of phase-matched periodograms averaged to construct the power in that segment, for different segment lengths between $0.5$ and $2.5$ seconds. The p-values indicate that the signal is most strongly detected for segments of $2\,\mathrm{s}$ duration, and there is an indication for intermittency in the p-values for that segment duration.}
\label{fig:rhessi_pvalues}
\end{center}
\end{figure}
\subsection{RHESSI}
\label{sec:rhessi_results}

\citealt{Watts06} searched segments of $t_{\mathrm{seg}} = 2.27$ seconds length, i.e. $1/3$ of the neutron star's rotational cycle, over a range of 19 successive cycles, starting $\sim 80 \, \mathrm{s}$ after the onset of the giant flare. They report the detection of a QPO at $626.5 \, \mathrm{Hz}$ with a significance of $6.6 \times 10^{-5}$, corrected for the number of trials, in this averaged periodogram when comparing to the theoretically expected distribution of powers for pure Poisson noise.

While previous studies constrained themselves to a single (arbitrary) segment length, in our re-analysis of the \rhessi\ data we varied the length of the segments between $t_{\mathrm{seg}} = 0.5 \, \mathrm{s}$ and $t_{\mathrm{seg}} = 2.5 \, \mathrm{s}$  in order to be sensitive to shorter signals, which may be buried in noise when taking the periodogram over too long a segment. This is not necessary for the \rxte\ data, since the signal is strong enough and the data is of high enough quality for the signal to be clearly observable even if it is considerably shorter than the segment length. For weaker signals and data of lower quality, a short signal can potentially be buried under the noise when looking at segments that are much longer compared to the duration of the QPO. 

\begin{figure*}[htbp]
\begin{center}
\includegraphics[width=1.1\textwidth]{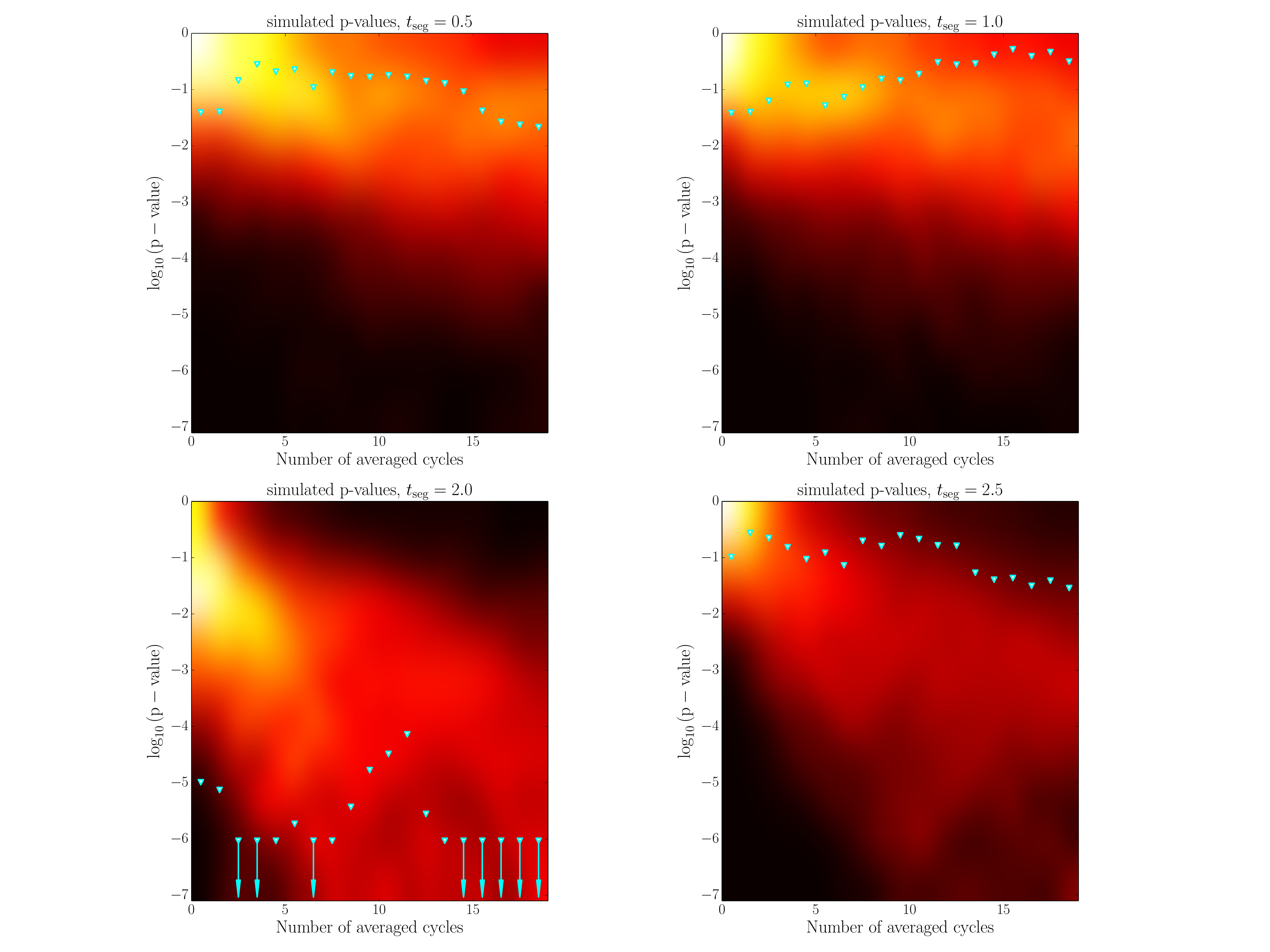}
\caption{\rhessi\ data: observed p-values versus $600$ simulations for different segment lengths. The colour map in the background corresponds encodes the p-value that simulated light curves with a QPO of $2$ second duration and a fractional rms amplitude of $0.1$ in every rotational cycle is likely to have as a function of the number of averaged cycles. We plot the p-values derived from the \rhessi\ data as light blue triangles, with arrows indicating upper limits from $10^{6}$ simulations. Simulated p-values below $10^{-6}$ are interpolations.}
\label{fig:rhessi_sims1_pvalues}
\end{center}
\end{figure*}
We subdivided each rotational cycle of the neutron star into $N_\mathrm{s} = 30$ segments, such that they start every $\Delta t = 7.5477/30 = 0.2534 \, \mathrm{s}$ apart, and overlap for $\delta t_\mathrm{seg} - 0.2534 \, \mathrm{s}$. Again, we use a higher number of segments per cycle to account for the poorer quality of the \rhessi\ data, and the fact we search shorter segments: for $15$ segments per cycle, the shortest segments will not overlap, and a signal split between two segments may not be detected at all.
For each segment we computed the periodogram, extracted the power at $626.5 \, \mathrm{Hz}$, and compared this power to those at the same frequency from segments of $N_{\mathrm{sim}}$ simulated light curves with the giant flare pulse profile, but smoothed out such that the QPO is removed. We varied the number of simulations $N_\mathrm{sims}$ for different segment lengths such that we could constrain the p-value robustly. However, we cut off our simulations at $N_\mathrm{sim} = 10^{6}$; larger runs are not computationally feasible. All p-values that fall below this value are quoted as upper limits.

The p-values for a simulated power in any segment to be higher than the power in any observed segment is shown in Figure \ref{fig:rhessi_pvalues}. If the signal is present intermittently, then the significance should decrease with increasing number of averaged cycles, since any additional cycle included in the average will only supply noise. On the other hand, if the signal is long-lived and persists over many cycles, then averaging more cycles should make the observed signal more significant, and the p-value should decrease with an increasing number of averaged cycles.

Interestingly, the highest significance for the QPO signal is not necessarily for an average of $19$ cycles, as reported in \citet{Watts06}, but there is a strong signal when averaging very few cycles: averaging $3$ or $4$ cycles results in a highly significant detection, whereas adding further cycles first decreases the significance, indicating that noise is being added. For averages of $12$ cycles or more, the p-value drops again. This could be an indication for intermittency of the QPO signal in the \rhessi\ data as well.

Because the signal-to-noise ratio is much lower for the \rhessi\ data than the \rxte\ data, we cannot repeat the analysis of Section \ref{sec:rxte_results}, where we searched for the presence of the $625 \, \mathrm{Hz}$ QPO in a single cycle to a very high phase resolution in order to determine its properties, on this data set. Instead, we simulate giant flare light curves from the original data set, with the $626.5 \, \mathrm{Hz}$ QPO smoothed out, and a signal at the same frequency injected back with varying parameters. We then compute p-values for the power at $626.5 \, \mathrm{Hz}$ in the same way as for the observed giant flare data, and compare these simulated p-values with those derived from the data. We test whether the data are consistent with two different hypotheses: (1) the QPO appears at the same phase in every rotational cycle, and is consistently present over all nineteen averaged cycles; (2) the signal is present in a small subset of cycles.

 \begin{figure*}[htbp]
\begin{center}
\includegraphics[width=1.1\textwidth]{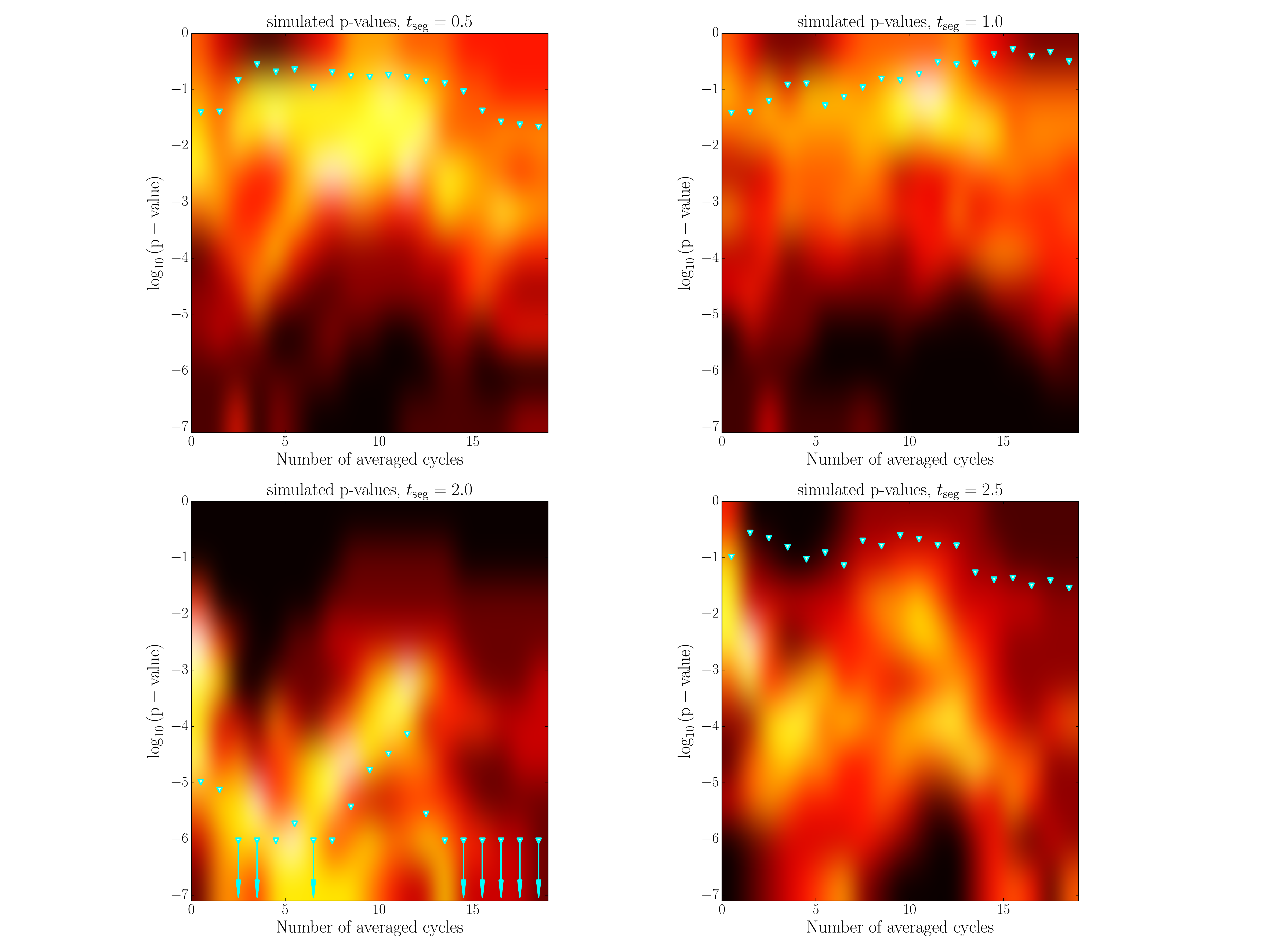}
\caption{\rhessi\ data: observed p-values versus $600$ simulations for different segment lengths. The colour map in the background corresponds encodes the p-value of simulated light curves with a QPO of $1$ second duration and a fractional rms amplitude of $0.4$ in the strongest cycle, as well as weaker sinusoidal signals in the preceding cycle as well as the second-to-last four cycles. We plot the p-values derived from the \rhessi\ data as light blue triangles, with arrows indicating upper limits from $10^{6}$ simulations. Simulated p-values below $10^{-6}$ are interpolations.}
\label{fig:rhessi_sims3_pvalues}
\end{center}
\end{figure*}

\subsection{Is the QPO present at the same phase in all cycles?}

In order to test the first hypothesis, we injected a sinusoidal signal at the same rotational phase ($2.07/(2\pi)$, i.e. the same rotational phase of the neutron star at which the QPO is observed in the segments with a $2\,\mathrm{s}$ duration) for all $19$ rotational cycles we searched. The start phase of the sinusoidal signal was randomised in each cycle, as well as in each simulated light curve. The resulting time series were randomised using a Poisson distribution to account for photon counting noise, and subjected to the same analysis procedure as the giant flare data to extract p-values as a function of the number of rotational cycles averaged, for four different time segment sizes.

In Figure \ref{fig:rhessi_sims1_pvalues} we show representative results for $600$ simulations with a QPO in every cycle of $2$ second duration and a fractional rms amplitude of $10\%$. Qualitatively, the simulations show similar p-values to the observed \rhessi\ data for the shorter segments, whereas for $2$-second segments, the simulations seem to underestimate the observed signal and fail to reproduce the trends of increasing and decreasing p-values as observed in the data. We note, however, that there is a considerable scatter in the p-values, especially when averaging many cycles: this indicates that for a given signal strength, realisations can differ widely. We have already shown this for the \rxte\ data in Figure \ref{fig:rxte_sims}, where there is a considerable scatter on the observed powers at $625 \, \mathrm{Hz}$ even for the strong signal in a single cycle, a problem that will be exacerbated by the lower data quality of the \rhessi\ data, as well as the data folding.

\subsection{Is the observed QPO with a signal present in a few cycles?}

Testing whether the QPO is only present in few cycles is not straightforward with the kind of forward-fitting technique employed here: a QPO could be present in any number of the 19 cycles considered here, and no potential QPO duration per rotational cycle or QPO amplitude can be excluded a priori. This leaves us with an enormous parameter space to traverse, while at the same time creating a large number of simulations and performing the same analysis as on the data for each possible parameter set becomes prohibitively computationally expensive. We thus restrict ourselves to few informed guesses to the possible distribution of QPOs, and with qualitative arguments for the simulations we considered. 

As in the preceding section, we removed any variability above $100 \,\mathrm{Hz}$ from the \rhessi\ giant flare light curve via smoothing, and added a sinusoidal signal at the original QPO frequency into a number of cycles.
We injected a strong, short signal (duration $1.0 \,\mathrm{s}$, fractional rms amplitude $0.4$) into the cycle where the highest power at $626.5 \,\mathrm{Hz}$ is observed in the \rhessi\ data, and a somewhat weaker sinusoidal signal (same duration, fractional rms amplitude $0.3$) in the preceding cycle.
Additionally, we introduced a longer, but weaker signal of $2\,\mathrm{s}$ duration and a fractional rms amplitude of $0.2$ into cycles $14$ to $18$, to mimic the downward trend of the p-values when averaging many cycles. We then simulated $600$ realisations from this model using a Poisson distribution to account for photon counting statistics in the detector.
This model qualitatively reproduces the trends observed in the p-values for all four segment lengths tested (see \ref{fig:rhessi_sims3_pvalues}), but overestimates the significance of detection for the longest segments searched. Compared with a model that includes a QPO in all 19 cycles, an intermittent QPO present in only a few cycles seems to be equally reasonable or favourable. The poor data quality leads to a large spread in p-values; it is thus difficult to draw strong conclusions from the data. However, given the p-values shown in Figure \ref{fig:rhessi_pvalues} and the outcome of the illustrative, but limited simulations performed here, there is no reason to prefer a long-lived signal over a short-lived, intermittent one.

\section{Discussion}
\label{sec:discussion}

The strongest conclusions we can draw come from a re-analysis of the \rxte\ data; the \rhessi\ data is of lower quality, and thus ambiguous. While the evidence for a short lifetime of the QPO in the \rxte\ data is fairly strong, the results from the \rhessi\ analysis could be interpreted either as a long-lived QPO or an intermittent one, and data quality is insufficient to reject either model.
 We have shown that the $625$Hz QPO is not present continuously throughout the $9$ cycles starting at $\sim\!\! 239 \, \mathrm{s}$ after the trigger at the low energies recorded with \rxte, as was inferred previously, and is instead concentrated within $\sim\!\! 2$ rotational cycles, during which it was excited and then decayed over the timescale
of $\sim\!\! 0.5$sec. While the origin of QPO excitation and re-excitation during the giant flare's tail is unknown, the
decay is expected on theoretical grounds. As was already discussed in the introduction, if the QPO reflects the oscillation of the $n=1$ crustal mode, it is expected to decay rapidly due to the crustal mode's strong coupling to the Alfven modes
in the magnetar's core. The calculations of e.g. \citet{vanHoven12}, find the timescale for this decay is $\sim\!\! 0.03$s (see, e.g., their Figure 11 where the transient nature of the decay is taken into account), which is more than an order of magnitude shorter than what is observed in our analysis. It remains to be seen how the difference between the theory and observations can be better reconciled. One of the possibilities is that so far 
all theoretical models have assumed that the magnetic field threads all of the core, and therefore the core provides a large reservoir for effectively absorbing the energy of the crustal motion. However, it may be that the magnetic field is concentrated in the outer parts of the neutron star, in which case the coupling is reduced. This may occur because the dynamo process that makes the field is most effective in
the outer layers, as suggested by \citet{bonanno2006}. Alternatively, if magnetars are born spinning rapidly, their subsequent spin-down drives the outward motion of the superfluid vortices in the core; this motion may effectively push the magnetic fields out of the core, due to strong interaction between the vortices and superconducting flux tubes \citep{ruderman1998,glampedakis2011}.

This re-analysis not only provides a fresh look into magnetar QPOs, but also demonstrates the power of model-oriented analyses in that context. The initial analyses carried out by \citet{Israel05,Strohmayer05,Strohmayer06,Watts06} were largely exploratory: QPO searches were carried out over large ranges of frequencies, time segments and numbers of cycles considered. As shown in this work, the potential parameter space for such searches is vast, and the sensitivity of a search is immediately and strongly limited by the number of alternatives considered. Any conclusions drawn from a search over a subset of these alternatives will be necessarily biased by the parameter choices made, and can thus potentially mislead a theoretical interpretation.

New approaches to the data, informed by hypotheses and questions posed by specific theoretical models, can overcome this problem. By testing specific model predictions, the data analysis can be made much more precise, and more informative with respect to the model predictions, even when the data quality is relatively poor. In this paper, by testing a specific prediction, we have shown that current data are compatible with current theoretical predictions of short decay times, even though this was not clear from the original, non-targeted, analysis (although it puts the emphasis back onto the question of excitation and re-excitation of oscillations). This is especially important in light of potential future giant flares observed with high-sensitivity instruments such as \fermi/GBM, which operates at similar energy ranges as \rhessi, but would provide data of unprecedented quality.

\acknowledgments
The authors thank Lucy Heil for useful discussions, and the referee for helpful suggestions. 
DH and ALW acknowledge support from a Netherlands Organization for Scientific Research (NWO) Vidi Fellowship (PI A. Watts).  
DH and YL acknowledge the Monash Reasearch Advancement grant (PI Yuri Levin) for partially funding collaborative visits
and research on magnetars.

\bibliography{bibliography}
\bibliographystyle{apj}

\end{document}